\documentclass{iaus}
\usepackage{graphicx}
\usepackage{lscape}

\title[Abundances in the disk-bulge connection] 
{Chemical abundances of planetary nebulae \\ in the disk-bulge connection}

\author[Roberto D.D. Costa, Oscar Cavichia \& Walter J. Maciel]   
{Roberto D.D. Costa$^1$, Oscar Cavichia$^1$ \and Walter J. Maciel$^1$}

\affiliation{$^1$IAG, University of S\~ao Paulo, 05508-090,
S\~ao Paulo/SP, Brazil \\ email: {\tt roberto@astro.iag.usp.br};
email: {\tt cavichia@astro.iag.usp.br}; \\ email: {\tt maciel@astro.iag.usp.br}}

\pubyear{2008}
\volume{254}  
\pagerange{1--5}
\jname{The Galaxy Disk in Cosmological Context}
\editors{J. Andersen, J. Bland-Hawthorn \& B. Nordstr\"om, eds.}
\begin{document}

\maketitle

\begin{abstract}
We report the spectrophotometric investigation of a planetary nebula sample located at the disk-bulge interface of the Milky Way. The main goal of this work was to determine the galactocentric distance where, according to the intermediate mass population, bulge and disk properties separate. In order to derive such distance, new abundances were derived for a sample of PN located at this region, and the results were combined with additional data from the literature. The abundance analysis indicates a chemical abundance distribution similar to that derived from bulge stars, as already pointed out by other authors. 

Statistical distance scales were then used to study the distribution of chemical abundances across the disk-bulge connection. A Kolmogorov-Smirnov test was used to find the distance in which the chemical properties of these regions better separate, resulting in a best value of 2.9 kpc to define the inner limit of the disk.

\end{abstract}

\section{Introduction}

Chemical properties of the galactic bulge and inner disk should intersect in a given
region. In this region bulge characteristics such as diversity in chemical abundances
should meet disk properties such as the inner edge of the radial gradient of chemical
abundances.

The main goal of this work is to separate intermediate mass populations from the bulge and inner disk. This was made through the analysis of the abundance distribution of planetary nebulae (PN) located in the inner Galaxy. New abundances for a PN sample were derived, then added to data from the literature for other objects. These results were combined and then statistically analyzed, in order to estimate a galactocentric distance where bulge and disk characteristics better separate. Ultimately, we intend to provide an additional constraint to chemical evolution models for the Galaxy.

\section{Sample and observations}

\begin{figure}[ht!]
\begin{center}
 \includegraphics[width=4in]{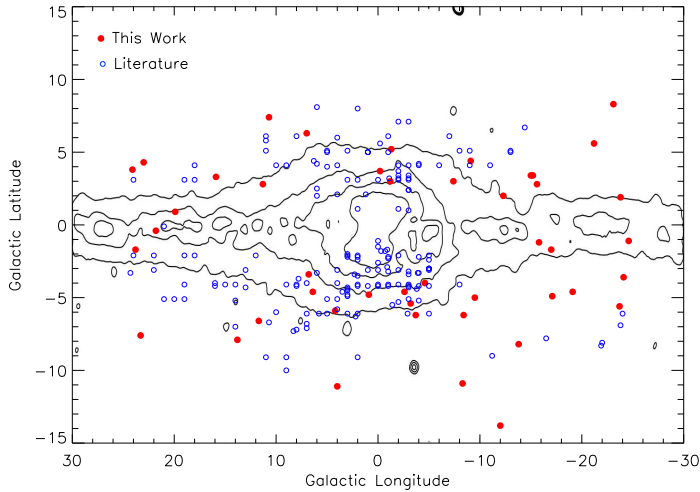} 
 \caption{Location of the PN sample.}
   \label{fig1}
\end{center}
\end{figure}

We selected 56 objects in the inner Galaxy based on their positions with respect to the 
galactic center ($|l|$, $|b|$), radio fluxes at 5 GHz and optical angular diameters ($\theta$), according to these criteria: i) $|l|<25^o$ and $|b|<10^o$; ii) F(5 GHz) $<$ 100 mJy; iii) $\theta <$ 12". Figure \ref{fig1} indicates the position of
 the objects for which new abundances were derived (in red), and those from the literature 
(in blue). The figure also shows the contours of the galactic bulge as seen from the infrared
observations of the COBE/DIRBE satellite (\cite[Weiland et al. 1994)]{Wei94}.

Observations were performed with the Brazilian 1.60 m telescope located at Pico dos Dias in southeast Brazil, with a Boller \& Chivens Cassegrain spectrograph. Each object was observed at least twice. For all objects we used a 1.5 arcsec width slit. Flux calibration was secured
through the observation of spectrophotometric standards (at least three each night). A 300 l/mm grating allowed a dispersion of 2.4 \AA /pixel. Details for the observations, derived
physical parameters and abundances for the sample are given by \cite[Cavichia et al. (2008)]{Cav08}.

\section{Results}
\subsection{Distance-independent correlations}

\begin{figure}[ht!]
\begin{center}
 \includegraphics[width=2.6in]{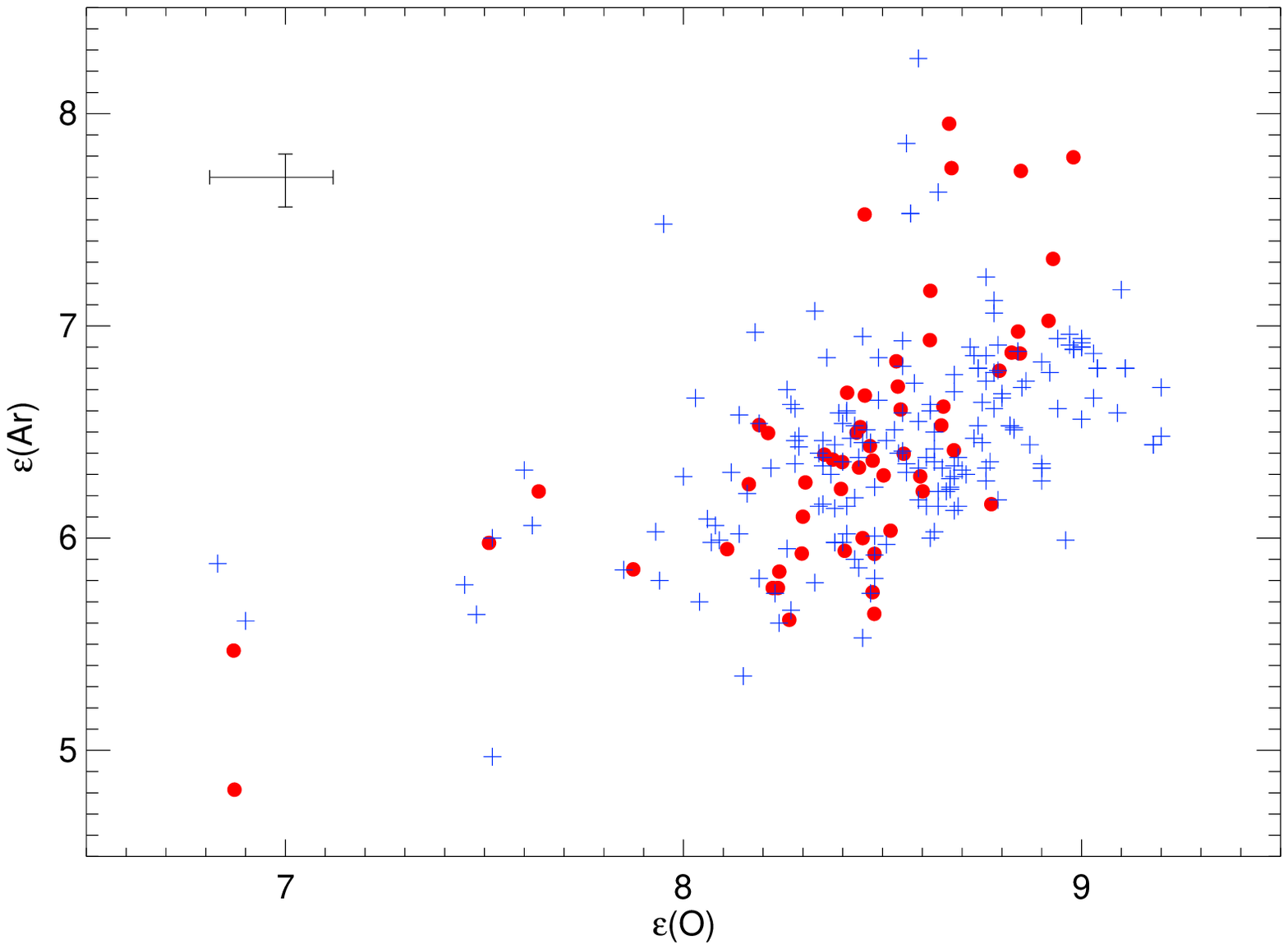} 
 \includegraphics[width=2.6in]{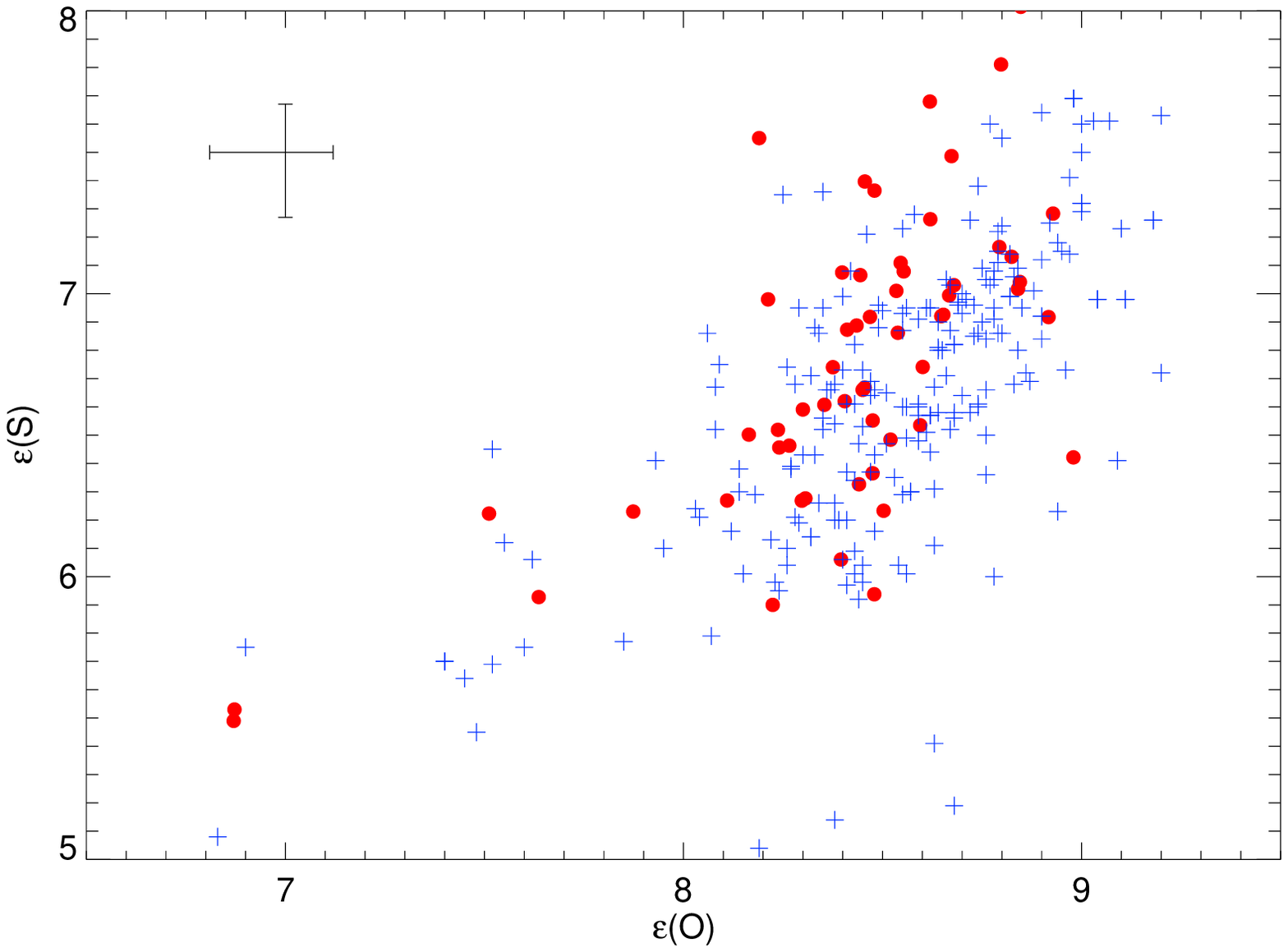}
 \caption{Distance-independent correlations between abundances of $\alpha$-elements.}
   \label{fig2}
\end{center}
\end{figure}

We used the derived chemical abundances to build distance-independent correlations between
$\alpha$-elements. Since they are produced basically by type-II supernovae, their abundances
reflect the chemical evolution of the Galaxy. These correlations are 
shown in Fig.\,\ref{fig2}. Mean errors for each element are given in the upper left 
corner of each graph.

It is easy to see that $\alpha$-element abundances are correlated, in the sense that older objects, formed in a poorer interstellar medium, present lower abundances of all these elements.
The same effect appears both for our sample and abundances collected from the literature. 
Therefore these values reflect the abundances of the interstellar medium at the progenitor
formation epoch, indicating the chemical evolution of the Galaxy.

\subsection{The bulge-disk interface}

\begin{figure}[bt]
\begin{center}
 \includegraphics[width=2.6in]{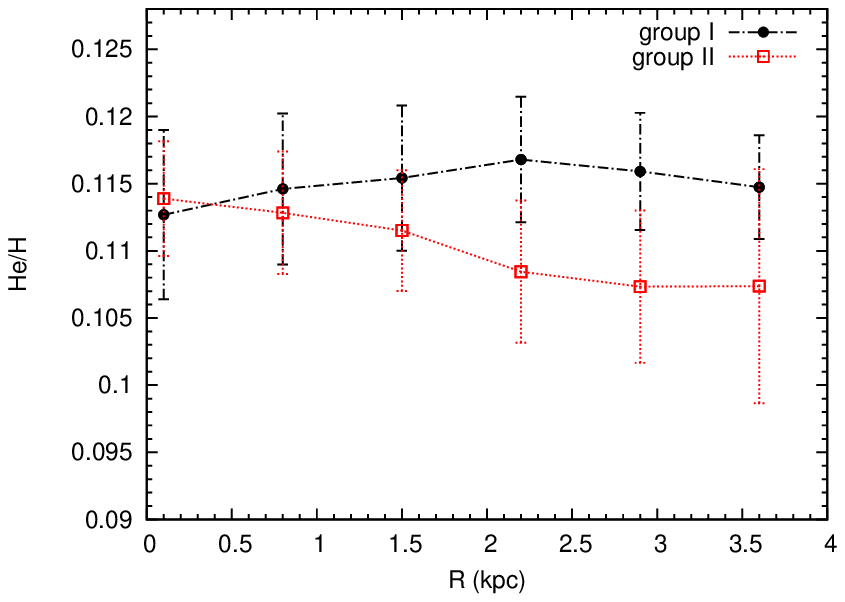} 
 \includegraphics[width=2.6in]{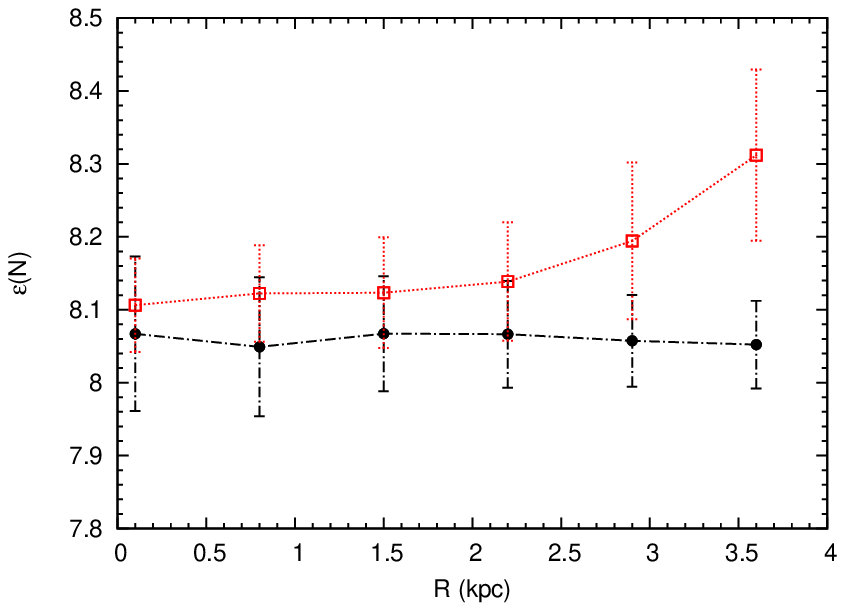}
 \includegraphics[width=2.6in]{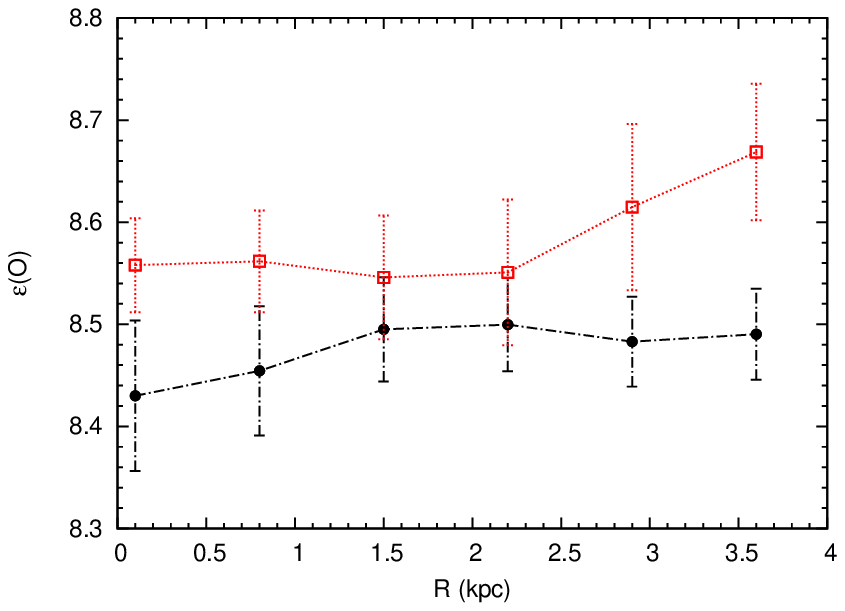}
 \includegraphics[width=2.6in]{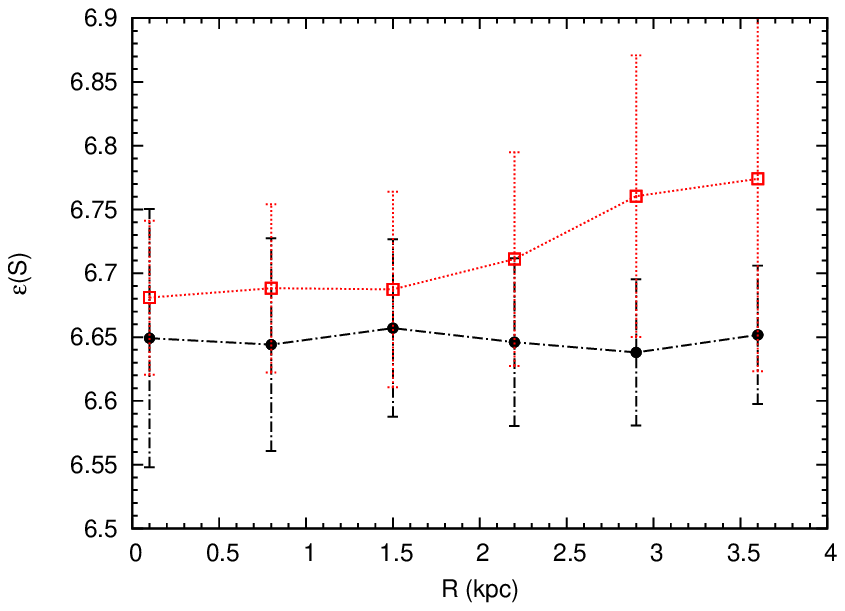}
 \includegraphics[width=2.6in]{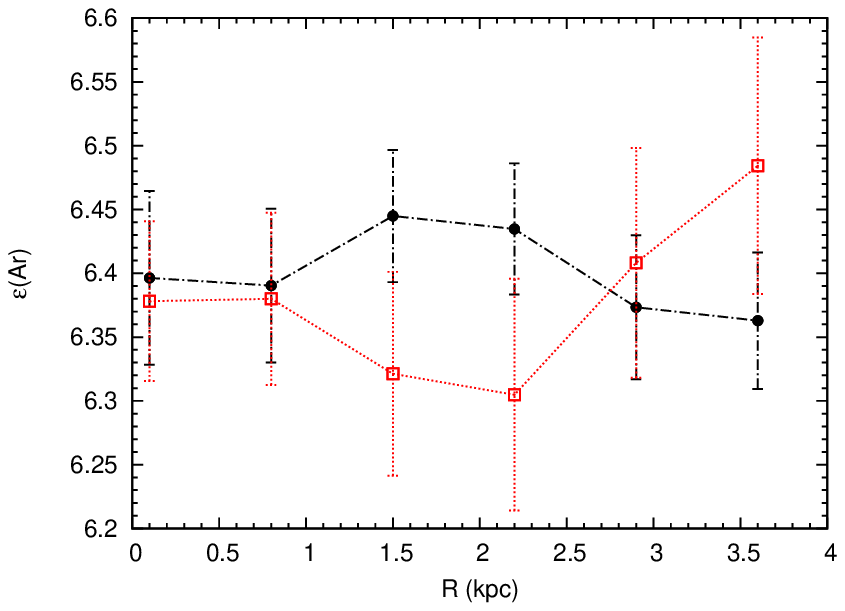} 
 \includegraphics[width=2.6in]{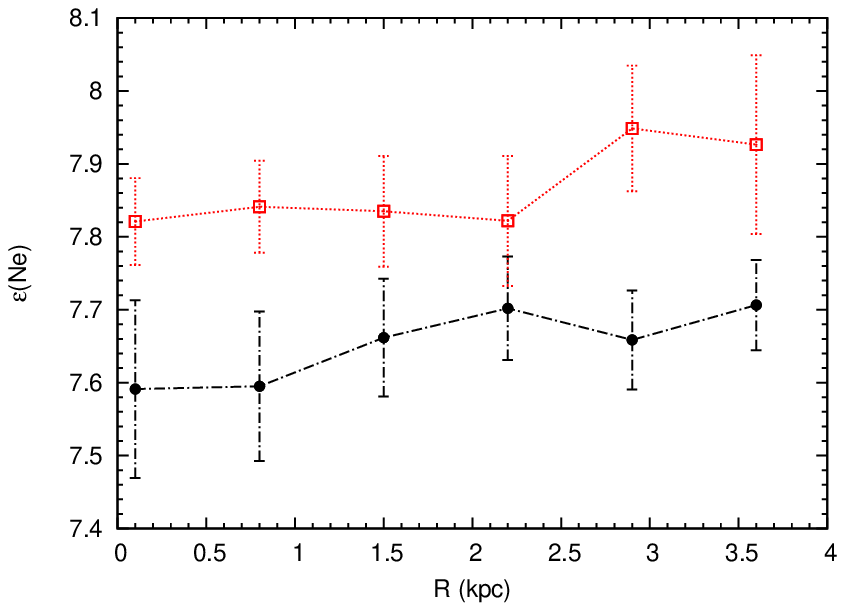}
 
\caption{Average abundances for different elements adopting the 
\cite[Cahn et al. (1992)]{Cahn92} distance scale, for different limits of the bulge-disk
interface (see text).}
   \label{fig3}
\end{center}
\end{figure}

\begin{figure}[ht!]
\begin{center}
 \includegraphics[width=2.5in]{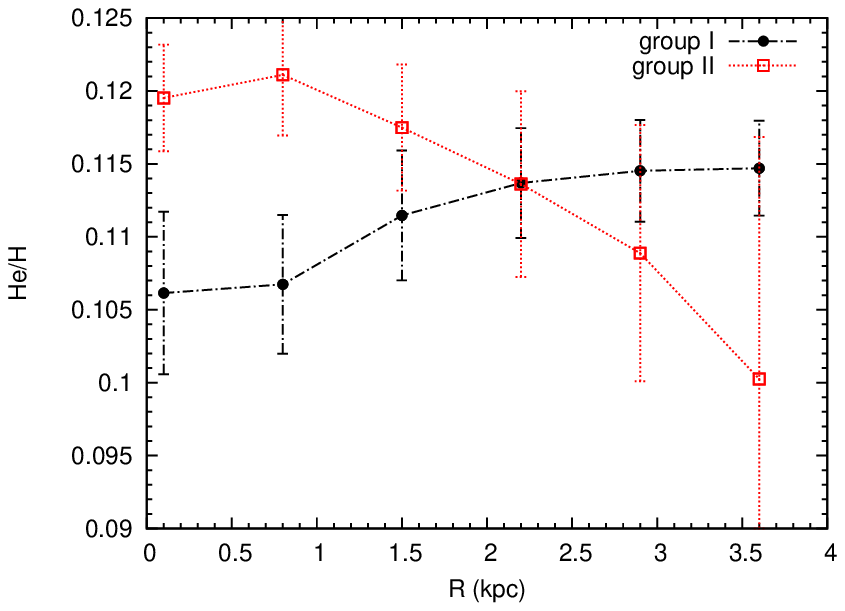} 
 \includegraphics[width=2.5in]{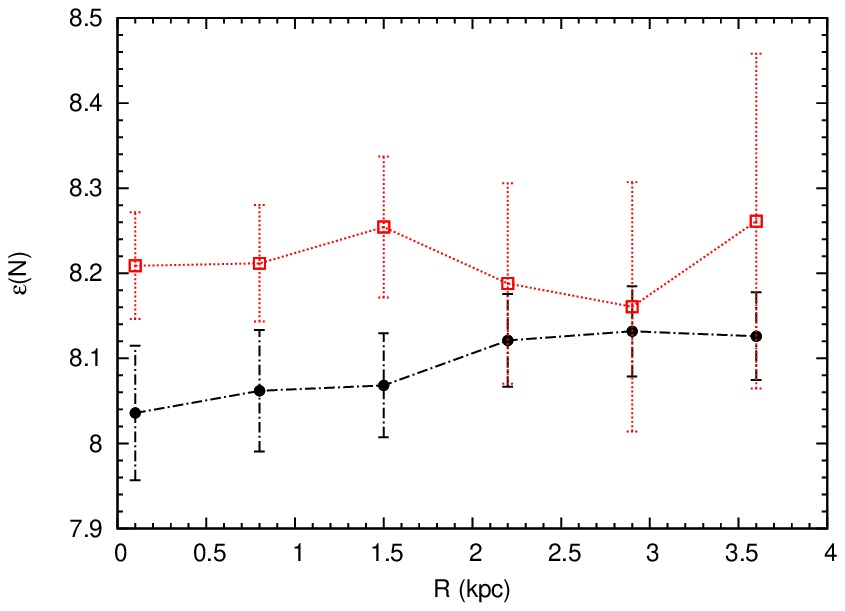}
 \includegraphics[width=2.5in]{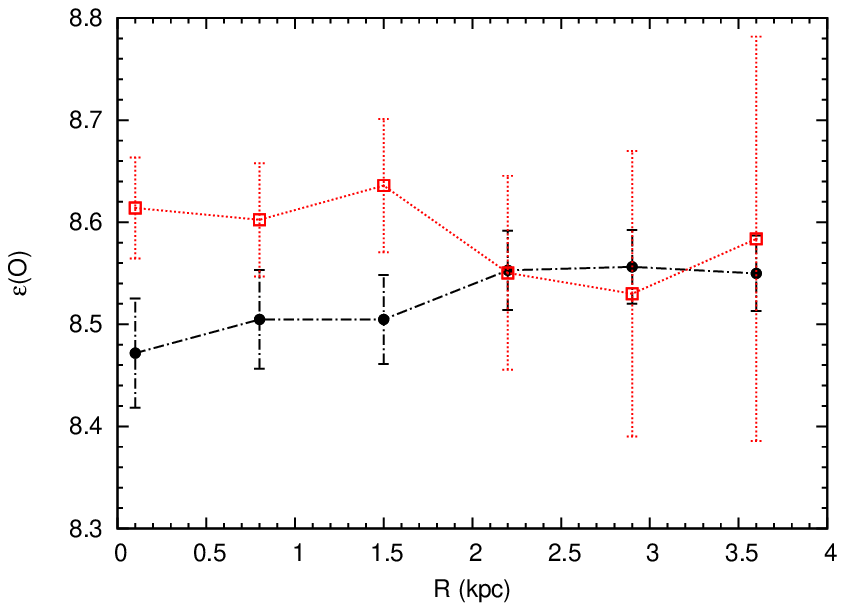}
 \includegraphics[width=2.5in]{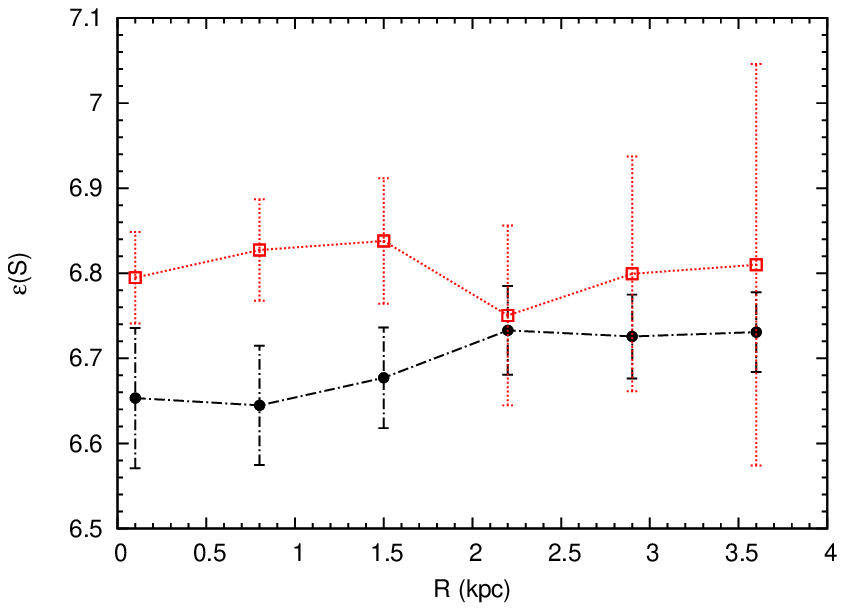}
 \includegraphics[width=2.5in]{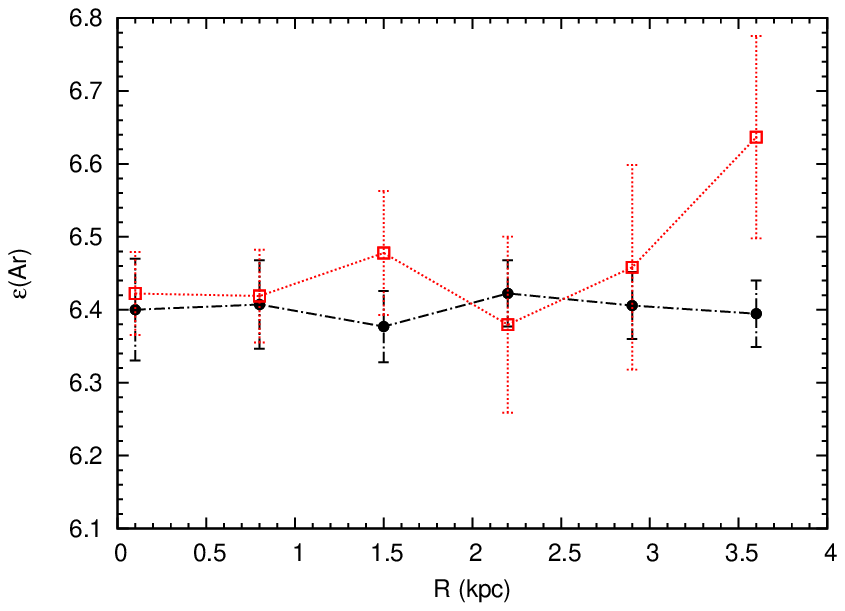}
 \includegraphics[width=2.5in]{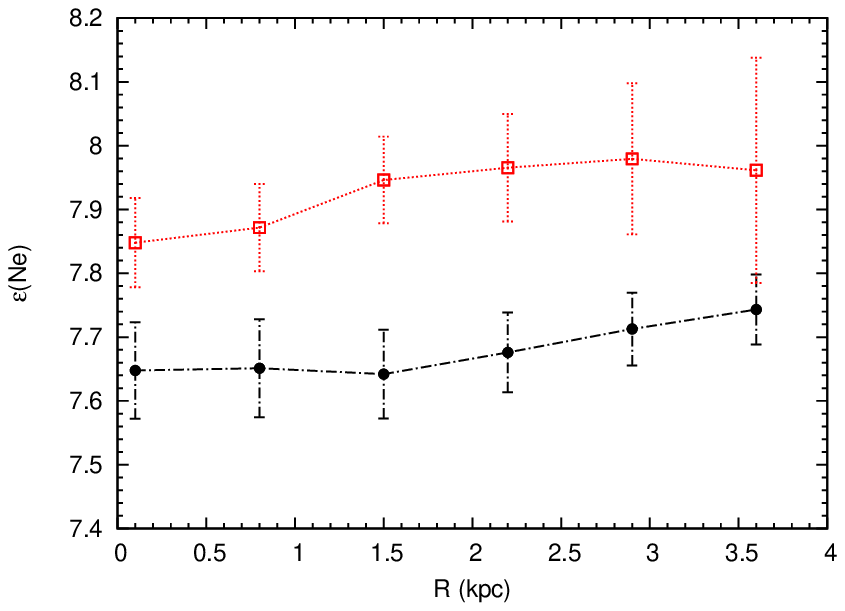}
 \caption{Average abundances for different elements adopting the 
\cite[Zhang (1995)]{Zhang95} distance scale, for different limits of the bulge-disk
interface (see text).}
   \label{fig4}
\end{center}
\end{figure}

One of the main goals of this work was to determine the galactocentric distance where bulge
and disk characteristics intersect. To estimate such a distance we adopted the following approach:

\begin{itemize}
\item{Statistical distances were selected for all objects in the sample using the statistical scales
of \cite[Zhang (1995)]{Zhang95} and \cite[Cahn et al. (1992)]{Cahn92}. Since these scales are
internally consistent but can present large differences for a given object, we chose to use
them separately, avoiding mixing distances from different scales.}

\item{Once the individual galactocentric distances were defined, we divided the sample into two
 distance groups: one with distances smaller than a given limit, which we called Group I, and other with distances larger than this limit, which we called Group II. For each group average abundances were calculated for the available elements. We considered
 limits from 0.1 to 3.6 kpc, in 0.7 kpc steps and, for each step, average abundances were
 calculated. The results are shown in Figs.\,\ref{fig3} and \ref{fig4}.}
 
\end{itemize}

Figure \ref{fig3} was built using the \cite[Cahn et al. (1992)]{Cahn92} distance scale, and 
Fig.\,\ref{fig4} using the \cite[Zhang (1995)]{Zhang95} scale. In each figure, Groups I and II
are represented respectively in black and red. Each pair of black/red points represents the average abundance for that element adopting a given limit for the bulge-disk
interface, and therefore each plot shows how the differences between both groups evolve adopting distinct limits to define them. Examining Fig.\,\ref{fig3} we see that for the 
$\alpha$-elements oxygen, sulfur and argon the differences between both samples reach a 
minimum between 2.5 and 3 kpc. For argon this effect is not clear, but already expected 
since the ionization correction factor used to derive its abundance can lead to 
uncertainties larger than those for other elements. 
For nitrogen and helium, whose abundances reflect the chemical
enrichment processes occurring during the evolution of the progenitor star, this minimum
difference at 2.5 kpc does not appear, as expected.

Figure \ref{fig4} shows the same plots using the \cite[Zhang (1995)]{Zhang95} scale. Examining
again the $\alpha$-elements, we see now that a minimum difference appears around 1.5 or 2 kpc,
depending on the element.

In order to ascertain a value for the galactocentric distance of the bulge-disk interface based
on the intermediate mass population represented by this planetary nebulae sample, we performed
a Kolmogorov-Smirnov test for Groups I and II at each step of the procedure described above. This test allowed to check which step in galactocentric distances separates more clearly 
both subsamples and resulted in a value of 2.9 kpc using the \cite[Cahn et al. (1992)]{Cahn92}
scale. 

Using the \cite[Zhang (1995)]{Zhang95} scale, the same test resulted in a better value of
1.5 kpc to the interface, however we adopted the 2.9 kpc value based on additional 
evidences that also point out to a distance around 3 kpc. According to \cite[Tiede \& Terndrup (1999)]{TT99}, the abundance
distribution in the inner Galaxy points to a discontinuity at 3 kpc. At the same distance
the gas density decreases rapidly (\cite[Portinari \& Chiosi 2000]{PC00}). The Galaxy evolution
models also point out to a severe decrease in the stellar formation rate for galactocentric
radii smaller than 3 kpc (\cite[Portinari \& Chiosi 1999]{PC99}). Combining these evidences with
our results, we can point out to a bulge disk interface for the intermediate mass population,
 marking therefore the outer border of the bulge and inner border of the disk, at 3 kpc. 

\begin{table}[ht!]
  \begin{center}
  \caption{Average abundances for bulge and inner disk objects.}
  \label{tab1}
 {\scriptsize
  \begin{tabular}{|l|c|c|}\hline 
{\bf Element} & {\bf Bulge} & {\bf Inner disk}  \\ \hline
   He/H & 0.116 $\pm$ 0.004 & 0.107 $\pm$ 0.006 \\
   $\varepsilon$(N) & 8.06 $\pm$ 0.06 & 8.19 $\pm$ 0.11 \\
   $\varepsilon$(O) & 8.48 $\pm$ 0.04 & 8.62 $\pm$ 0.08 \\
   $\varepsilon$(S) & 6.64 $\pm$ 0.06 & 6.76 $\pm$ 0.11 \\
   $\varepsilon$(Ar) & 6.37 $\pm$ 0.06 & 6.41 $\pm$ 0.09 \\
   $\varepsilon$(Ne) & 7.66 $\pm$ 0.07 & 7.93 $\pm$ 0.09 \\ \hline
  \end{tabular}
  }
 \end{center}
\end{table}

 \newpage
 
 Finally, Table \ref{tab1} shows the average abundances for bulge and inner disk objects, 
 adopting the 2.9 kpc interface. These abundances are similar, taking into account the large
dispersion found in the bulge, but it can be seen that the bulge
abundances of $\alpha$-elements are systematically lower than the inner disk
abundances, so that the separation between the two populations can
be clearly defined.


\begin{thebibliography}{}

\bibitem[Cavichia et al. (2008)]{Cav08}
{Cavichia, O., Costa, R.D.D. \& Maciel, W.J.} 2008, in preparation

\bibitem[Cahn et al. (1992)]{Cahn92}
{Cahn, J.H., Kaler, J. \& Stanghellini, L.} 1992
\textit{A\& AS}, 94, 399

\bibitem[Portinari \& Chiosi (1999)]{PC99}
{Portinari, L. \& Chiosi, C.} 1999
\textit{A\& A} 350, 827

\bibitem[Portinari \& Chiosi (2000)]{PC00}
{Portinari, L. \& Chiosi, C.} 2000
\textit{A\& A} 355, 929

\bibitem[Tiede \& Terndrup (1999)]{TT99}
{Tiede, G.P. \& Terndrup, D.M.} 1999
\textit{AJ} 118, 895

\bibitem[Weiland et al. (1994)]{Wei94} 
{Weiland, J. L., et al.} 1994
\textit{ApJ} 425, L81

\bibitem[Zhang, C.Y. (1995)]{Zhang95}
{Zhang, C.Y.} 1995
\textit{ApJS}, 98, 659

\end{thebibliography}
\end{document}